\documentstyle[11pt,newpasp,twoside,epsf]{article}
\def\edcomment#1{\iffalse\marginpar{\raggedright\sl#1\/}\else\relax\fi}
\reversemarginpar

\begin{document}
\title{Constraints on Intrinsic UV Absorption in NGC 3783}
\author{Jack R. Gabel \& Steven B. Kraemer}
\affil{Catholic University of America and Laboratory for Astronomy and
Solar Physics, NASA's Goddard Space Flight Center, Code 681, Greenbelt,
MD 20771}
\author{D. Michael Crenshaw}
\affil{Department of Physics and Astronomy, Georgia State University,
Atlanta, GA 30303}

\begin{abstract}
   Results from an analysis of the intrinsic UV absorption in the Seyfert 1 galaxy NGC 3783
are presented. We focus on two new results that demonstrate techniques for deriving important
constraints on the physical conditions and geometry of the absorbers.
First, using variability in the spectrum, the emission-line profile is separated into distinct 
kinematic components and the effect on the interpretation of covering factors and column densities
is demonstrated. Second, measurements of the 2s2p $^3$P metastable levels of C$^{+2}$ derived
from the C III*1175-76 absorption multiplet are presented.  New calculations
of the metastable level populations are given and shown to provide
a powerful diagnostic of the density (and thus location) and temperature in an absorber.
\end{abstract}

\vspace{-0.2cm}

\section{Introduction}
    Recent surveys with high-resolution UV spectra have shown that 
mass outflow is common in Seyfert 1 galaxies, appearing as blueshifted absorption in more than
half of observed objects (Crenshaw et al. 1999).
Detailed studies are needed to determine the physical conditions 
(density, ionization structure, total gas column) and geometry of the absorbers. 
Here, we focus on constraints derived on the absorbers in the UV spectrum of NGC 3783 to demonstrate
some techniques that can be used to probe mass outflow in AGNs.      
\section{Isolating Emission-Line Components: Effect on Covering Factor}
   Since the background AGN light that the absorption features are imprinted on is
comprised of distinct emission sources with different flux distributions,
sizes, and geometries, the most general treatment of absorption features should take into account
the different covering factors associated with each source.
The Lyman series lines in the averaged STIS and {\it FUSE}
spectra were used to separate the individual covering factors of the continuum and emission
line sources in NGC 3783 (Gabel et al. 2003). 
By incorporating variability of the emission-line profiles, that analysis can be extended
to treat {\it distinct emission-line regions}.  Figure 1 shows the C IV, Ly$\alpha$, and N V 
emission-line profiles in both high-state
(solid line) and low-state (dotted line) STIS spectra.
The low-state profiles have been scaled to match the flux
in the high velocity wings of the high-state spectrum.
The profiles diverge near line center due to the superposition of a varying broad line region component and a non-varying
narrower component (an intermediate line region, ILR).  
The ILR flux profiles (dashed lines) were solved as described in Gabel et al. (2004, in preparation).
\begin{figure}[ht]
\plotfiddle{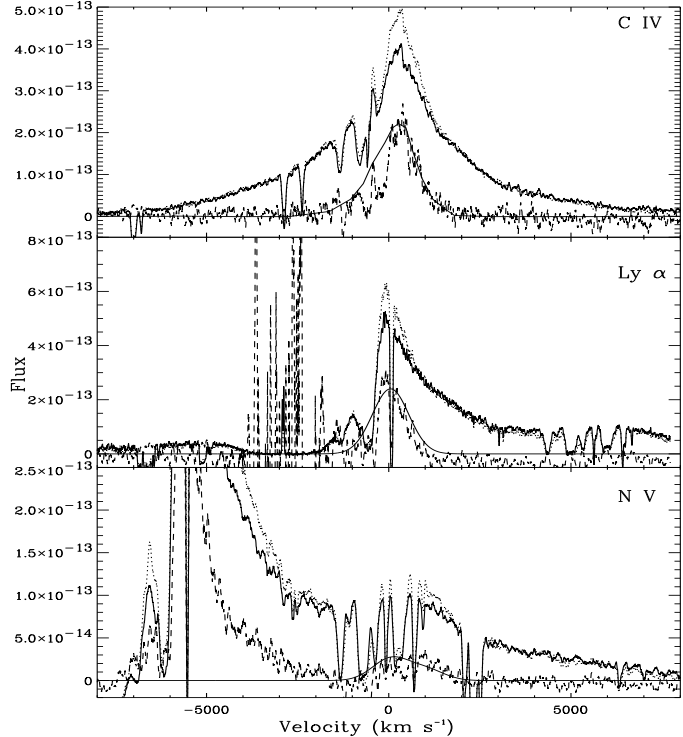}{3.45in}{0}{50}{40}{-150}{-33}
\caption{Emission line region profiles of averaged high-state and low-state
spectra for C IV, Ly$\alpha$, and N V. Low state profiles (dotted lines) are
scaled to match the high-state fluxes in the high-velocity wings. The dashed lines
give the non-varying (ILR) component.}
\end{figure}

    Figure 2 shows the N V absorption profiles for two normalizations to demonstrate the effect
of treating the covering factors of distinct emission components separately.
In the left panel, the absorption has been normalized simply by
dividing the total observed spectrum by our fit to the continuum plus total emission-line flux, 
i.e., the BLR + ILR. In the right panel, the ILR flux has first been subtracted out of both the data 
and emission-line fits, which
is equivalent to assuming the ILR is completely unocculted by the absorbers.
For each normalization, both the high-state (top panel) and low-state (bottom panel) spectra are
plotted for comparison and the two doublet lines are overlayed, shifted to coincide in radial 
velocity.  This shows that if the covering factors of the emission components are assumed
to be the same, then the N V absorption is not saturated and has effective covering factors
in the range $\approx$ 0.6 -- 0.65 for components 1 -- 3.
The similarity in the absorption depths between the high and low-states (for both
doublet lines) in the left panel implies there is little change in the column densities for this scenario.
Conversely, if the ILR is unocculted, the equivalent absorption depths of the two doublet lines 
indicates the absorption is saturated in both the
low and high states in some of the components.
This leads to significant differences in interpreting the physical conditions and
variability in the absorption. 

\begin{figure}
\plottwo{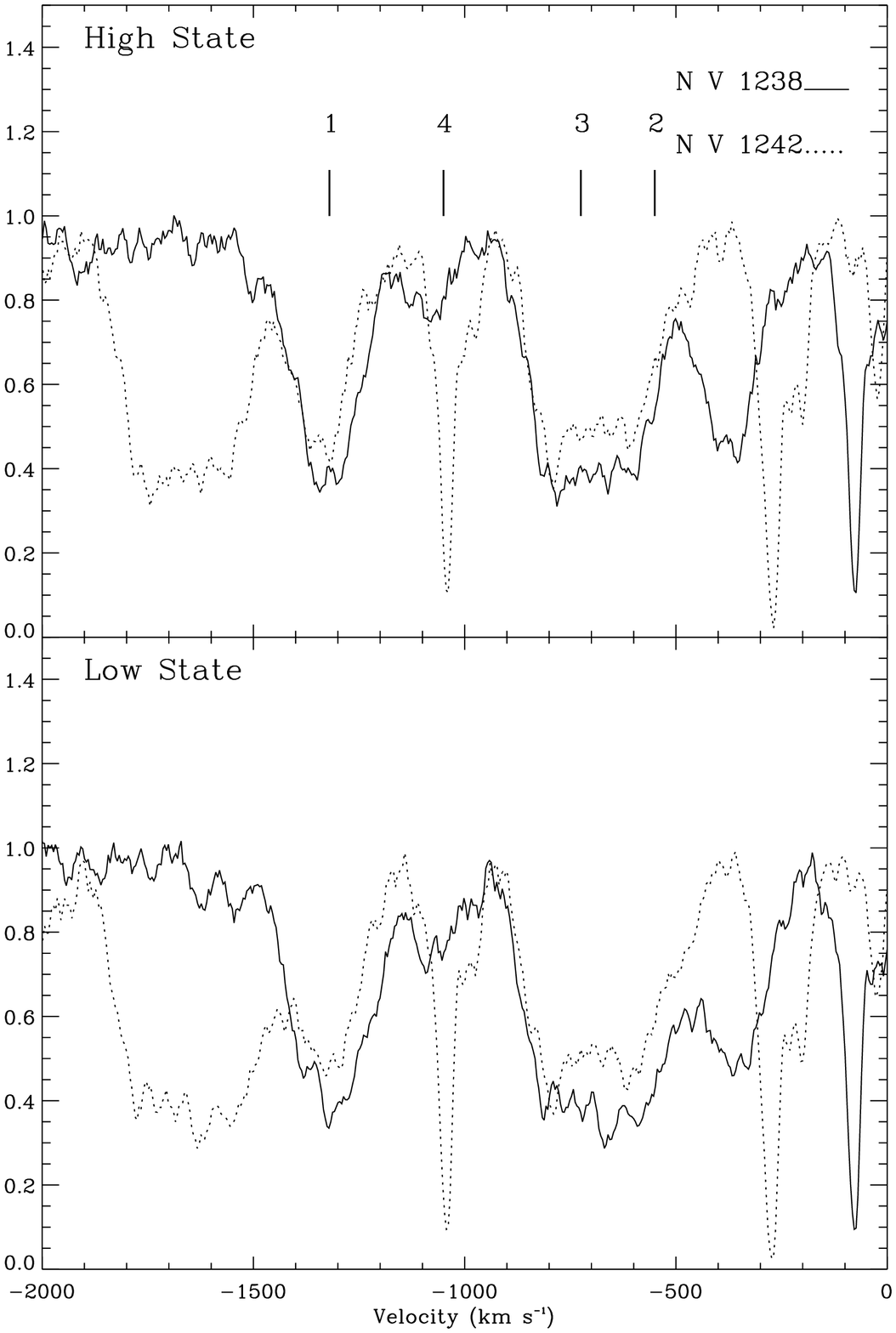}{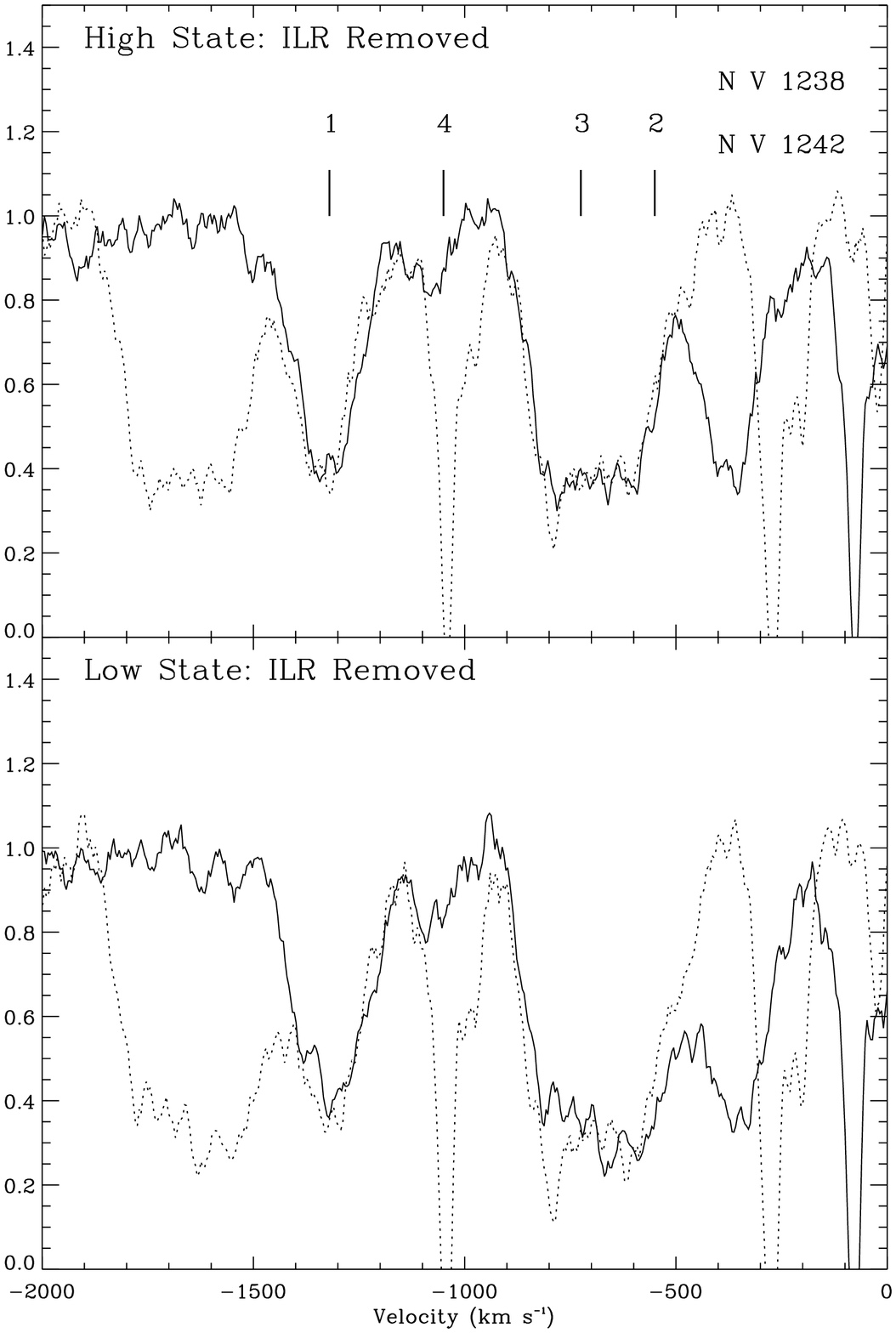}
\vspace{0.1cm}
\caption{Effect of treating individual covering factors of distinct emission components.
N V absorption profiles shown for two normalizations: dividing by fit to total emission (left panel)
and first subtracting out the ILR profile (right panel). The high (top) and low (bottom) state
spectra are shown for each case, and the N V 1242 line (dotted) is overlayed on the N V 1238 line.}
\end{figure}

\vspace{-0.1cm}

\section{C III* $\lambda$1175 Absorption as a Density Diagnostic}

    The C III* $\lambda\lambda$1175 multiplet lines have been used as a 
density diagnostic for AGN absorbers in several studies. However, as pointed
out by Behar et al. (2003, ApJ, in press), the
high densities derived in these studies (e.g., Gabel et al. 2003) 
were based on calculations of level populations that only treated the
$^3$P$_1$ level.
The $J=$0 and 2 levels have much lower radiative transition probabilities
to the ground state and thus are populated at
densities that are lower by several orders of magnitude (Bhatia \& Kastner 1993).
We have computed the relative populations of the $^3$P$_J$ levels, extending the
results of Bhatia \& Kastner (1993) down to electron temperatures expected
for the photoionized UV absorbers seen in AGNs.
Collisional excitation and de-excitation and radiative decay
between the six lowest terms/levels of the $C^{+2}$ ion were included in our calculations. 
\begin{figure}[ht]
\plotfiddle{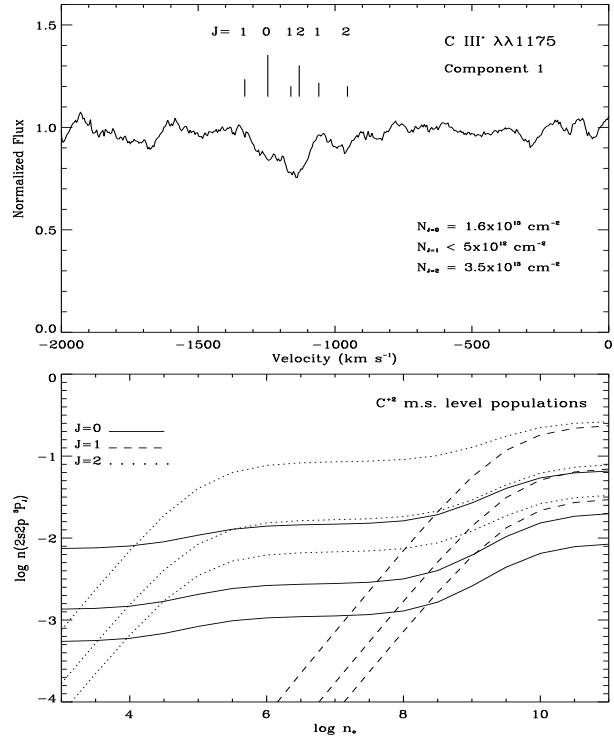}{3.4in}{0}{50}{40}{-160}{-32}
\caption{C III*1175 absorption showing the resolved mutliplet lines (top panel). Column densities
measured for each metastable level are given below the spectrum.  Calculation of C$^{+2}$ metastable
level populations (bottom panel) as a function of density for $T_e$=16000(lower), 20000, and 40000(upper) K.}
\vspace{-0.2cm}
\end{figure}
The top panel of Figure 3 shows the C III*1175 absorption complex in NGC 3783, with the location of the six multiplet
lines marked and identified by the $J$ level of the transition. Measured column densities for each 
metastable level are given below the spectrum.
The bottom panel shows the computed populations of the metastable levels over a large range in
density for $T_e=$ 16000, 20000, and 40000 K (bottom to top).  
Thus, if the absorption is sufficiently narrow to resolve and measure individual lines in the C III* multiplet,
the {\it ratios} of the excited level populations give a very tight constraint on the electron density.
For NGC 3783, the $J=$2 : $J=$0 column density ratios, $N_{J=2} / N_{J=0} =$ 2.2,
gives $n_e$=3$\times$10$^4$~cm$^{-3}$ , which is largely insensitive to temperature.
Combining this with a measurement of the ground state population provides a stringent temperature diagnostic.
The full implications of this measurement for the absorbers in NGC 3783 will be given in Gabel et al. (2004,
in preparation).

\end{document}